\documentclass{iaus}
\usepackage{graphicx}

\title[The Sun in Time] 
{The Sun in Time: Age, Rotation, and Magnetic Activity of the Sun and Solar-type Stars and Effects on Hosted Planets}

\author[E.F. Guinan \& S.G. Engle]   
{Edward F. Guinan$^1$ \& Scott G. Engle$^{1,2}$}

\affiliation{$^1$Department of Astronomy \& Astrophysics, Villanova University,
\break Villanova, PA 19085, USA
\break email: edward.guinan@villanova.edu
\break$^2$Centre for Astronomy, James Cook University,
\break Townsville, QLD 4811, Australia
\break email: scott.engle@villanova.edu}

\pubyear{2009}
\volume{258}  
\pagerange{}
\date{?? and in revised form ??}
\jname{The Ages of Stars}
\editors{E.E. Mamajek \& D. Soderblom, eds.}
\begin{document}

\maketitle

\begin{abstract}
Multi-wavelength studies of solar analogs (G0--5~V stars) with ages from $\sim$50 Myr to 9 Gyr have been carried out as part of the ``Sun in Time'' program for nearly 20 yrs. From these studies it is inferred that the young (ZAMS) Sun was rotating more than 10$\times$ faster than today.  As a consequence, young solar-type stars and the early Sun have vigorous magnetohydrodynamic (MHD) dynamos and correspondingly strong coronal X-ray and transition region / chromospheric FUV--UV emissions (up to several hundred times stronger than the present Sun). Also, rotational modulated, low amplitude light variations of young solar analogs indicate the presence of large starspot regions covering $\sim$5--30\% of their surfaces.  To ensure continuity and homogeneity for this program, we use a restricted sample of G0--5~V stars with masses, radii, T$_{\rm eff}$, and internal structure (i.e. outer convective zones) closely matching those of the Sun.  From these analogs we have determined reliable rotation-age-activity relations and X-ray--UV (XUV) spectral irradiances for the Sun (or any solar-type star) over time. These XUV irradiance measures serve as input data for investigating the photo-ionization and photo-chemical effects of the young, active Sun on the paleo-planetary atmospheres and environments of solar system planets.  These measures are also important to study the effects of these high energy emissions on the numerous exoplanets hosted by solar-type stars of different ages. Recently we have extended the study to include lower mass, main-sequence (dwarf) dK and dM stars to determine relationships among their rotation spin-down rates and coronal and chromospheric emissions as a function of mass and age.  From rotation-age-activity relations we can determine reliable ages for main-sequence G, K, M field stars and, subsequently, their hosted planets.  Also inferred are the present and the past XUV irradiance and plasma flux exposures that these planets have endured and the suitability of the hosted planets to develop and sustain life.

\end{abstract}

\firstsection 
\section{Introduction and Background on the Evolution of the Sun}

Changes in the Sun's luminosity, temperature and radius over its $\sim$10 Gyr main-sequence lifetime, arising from nuclear evolution, are explicable (except for ``minor'' details) by modern stellar evolution theory and models (Basu et al. 2000). However, an extrapolation of the Sun's magnetic dynamo-related properties back in time to study their manifestations -- such as chromospheric and coronal emissions, flares and coronal mass ejections, winds, and sunspots -- cannot be reliably modeled or predicted by theory alone.  Even for the present Sun, the solar dynamo and magnetohydrodynamic (MHD) dynamo, and resulting models, have problems explaining the observed magnetic-related phenomena -- such as the $\sim$11/22 year sunspot/magnetic cycle and the magnetic heating of the Sun's outer atmosphere.  But much progress in solar and stellar dynamo theory is being made (see e.g. Dikpati 2005).  However, we can trace the Sun's magnetic past by utilizing carefully selected solar analogs (with different ages and rotations) that serve as proxies for the Sun over its main-sequence lifetime.  To address this problem, we established the ``Sun in Time'' program nearly 20 years ago.  Unlike most other studies of stellar rotation-age-activity, this study focuses specifically on stars with nearly identical basic physical properties to the Sun.  Moreover, the program aims at determining changes in the coronal/chromospheric X-ray to UV (XUV) fluxes over the Sun's main-sequence lifetime and the effects that these high energy emissions have on hosted planets.  Related to the theme of this symposium, we also have developed rotation-age-activity relations for solar-type stars that permit the age of solar-like field stars to be determined from their rotation periods or from calibrated magnetic activity indicators such as Ca {\sc ii} $H \& K$ chromospheric emission (R'$_{HK}$), X-ray luminosity (L$_{\rm x}$) and spot-modulated brightness variations.
  
Studies of single main-sequence G and K-type stars in open clusters and stellar moving groups with known ages show that stars rotate more slowly with age and have corresponding decreases in magnetic activity such as chromospheric Ca {\sc ii} $HK$ emissions (e.g. Skumanich 1972; Soderblom 1982, 1983).  The observed decrease in rotation with age (for solar-type and cooler stars) results from angular momentum loss due to magnetically threaded winds (see Schrijver \& Zwann 2000 and references therein).  The Sun in Time project was motivated, in part, by the early studies of rotation, age \& activity of cool stars by Soderblom (see e.g. Soderblom 1982, 1983).  Some recent examples of related studies are given in Barnes (2007), Mamajek \& Hillenbrand (2008), several papers given in this volume (e.g. Barnes, Meiborn \& Mamajek) and the numerous references given in these papers.   

\section{The Sun in Time Program}

As an integral part of the Sun in Time program, studies of solar analogs (G0--5~V stars) have been carried out across the electromagnetic spectrum (see Dorren \& Guinan 1994; G\"udel et al. 1997; Guinan et al. 2003; Ribas et al. 2005).  The study utilizes X-ray, EUV, FUV \& UV data secured by us and others (including archival observations) from NASA, ESA and Japanese space missions as well as ground-based photometry and spectroscopy.  These stars serve as proxies for the Sun (and other solar-mass stars) and their ages essentially cover the main-sequence lifetime of the Sun (from $\sim$50 Myr to 9 Gyr).  From these studies (and others) it is inferred that the young (ZAMS) Sun was rotating more than 10$\times$ faster than today. See Fig. 1a for a plot of 1/P (days$^{-1}$) versus age for representative program stars.  This plot includes both G- and dM-type stars, so the behavior of the two stellar types can be seen and compared.  As indicated in the figure (and discussed later in Section 7), dM stars rotate more slowly than corresponding G-type stars of similar ages. As a consequence of rapid rotation, young solar-type stars and the young Sun (as well as dM-type stars) have vigorous magnetic dynamos and correspondingly strong coronal (X-ray) and transition region / chromospheric (FUV--UV) emissions up to several hundred times stronger than the present Sun.  Also, the rotationally modulated, low amplitude light variations of young solar analogs indicate the presence of large starspot regions that cover $\sim$5--30\% of their surfaces.  As an example, the coronal X-ray (0.2--2.5 keV) luminosities (L$_{\rm x}$) of the program stars are plotted versus age (Fig. 1b).  The least-squares fits to these data are provided within the figure.  Note the large decrease in L$_{\rm x}$ (by $>$ 1000$\times$) between young and older solar-type stars. 

To ensure consistency and homogeneity, the Sun in Time program uses a restricted sample of G0--5~V stars with masses, radii, T$_{\rm eff}$ and internal structure (i.e. outer convective zones) closely matching those of the Sun.  The only major ``free parameters'' are age and rotation.  From these analogs we have determined reliable rotation-age-activity relations and XUV spectral irradiances (flux/cm$^2$/sec at a reference distance of 1 AU) for the Sun over time.  These XUV irradiance determinations serve as inputs to investigate the photo-ionization and photo-chemical effects of the young, active Sun on paleo-planetary atmospheres and environments.  These measures are also important to study the numerous exoplanets hosted by solar-type stars of different ages.  Recently we have extended the study to include lower mass, main-sequence (dwarf) dK and dM stars to determine their spin-down rates and coronal and chromospheric emissions as a function of mass and age. The present and past XUV irradiances that these planets experience can then be estimated and the suitability of the hosted planets to retain their atmospheres and develop/sustain life can be assessed.  Also, from the studies by Wood et al. (2002, 2005) of solar-type, stellar astrospheres (equivalent to the solar heliosphere) using Hubble Space Telescope (HST)  H {\sc i} Lyman-$\alpha$ spectroscopy, it appears that the young Sun had more dense stellar winds and plasma outflows back to at least $\sim$3.5 Gyr ago.  As discussed by Lammer et al. (2003, 2008) and others, the interplay of strong XUV ionizing radiation and high plasma fluxes (winds), inferred for the young Sun, played a major role in the erosion of juvenile planetary atmospheres and loss of water for Venus and Mars.  More details on this work are given later.  

\begin{figure}
\includegraphics[height=2in,width=5.3in,angle=0]{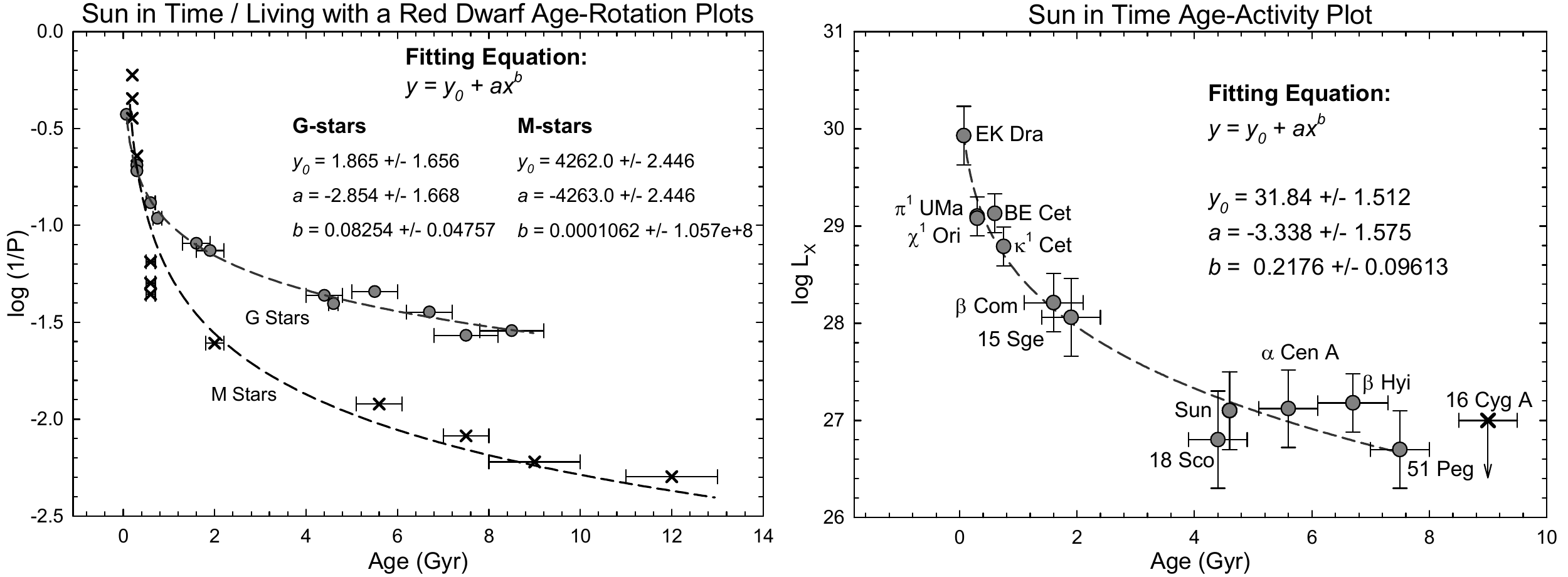}
\caption{{\bf Left (a) --} shows a plot of 1/P$_{\rm rot}$ (angular velocity $\Omega$ $\propto$ 1/P$_{\rm rot}$) of the solar-type stars in units of 1/days plotted against stellar age.  Also plotted for comparison are the same quantities for a representative sample of dM0--5 stars.  As shown in the figure, both the G- and dM-stars decrease in angular velocity with age. For a given age the dM stars have longer rotation periods indicating that they spin down faster than the more massive G stars.  As discussed in Section 7, the rapid magnetic braking of dM stars is most likely due to more efficient dynamos (from their deeper CZs) and their lower masses. {\bf Right (b) --} shows a plot of the mean $\log$ L$_{\rm x}$ (ergs/s) versus age for the solar-type stars.  As shown in the plot, the mean X-ray luminosity of the solar analogs decreases by 10$^3$ times from the youngest to solar-age stars in the sample.  The least squares fitting equations for both plots are given within the figures.}\label{fig:rotlx}
\end{figure}

The Sun in Time program utilizes a relatively small number of stars ($\sim$25 primary stars and about $\sim$30 secondary stars with less certain properties or measurements) that have similar basic physical properties such as mass (or mass parameterized by spectral types and colors). The homogeneity of the sample of $\sim$25 stars (except for age and rotation) ensures that the depth of the outer convective zone (CZ) -- which is important for a solar/stellar dynamo -- is nearly identical for all stars.  This is essential because the CZ, together with rotation, plays a crucial role in the magnetic dynamo, the generation of magnetic fields and the resulting outer atmospheric heating of solar-type (and cooler) stars -- which, in turn, impacts the properties of the chromospheres, Transition Regions (TRs), coronae, flares and coronal winds. The program also utilizes rotation periods, instead of $v \sin i$ measures, to serve as the angular velocity ($\Omega$ $\propto$ 1/P$_{\rm rot}$).  Most of the program stars are nearby, bright and have accurate distances. Thus they are accessible for securing measures of coronal X-ray \& EUV fluxes, and accurate measures of transition region and chromospheric emission fluxes. Many of the target stars now have long-term photometry, permitting the investigation of activity cycles and differential rotation.  Also, from the multi-band photometry, the properties of the starspots can be determined -- such as spot temperatures, areal coverage, surface distributions and differential rotation (see Messina et al. 2006 and references therein). The primary science objectives are given below.

\subsection{Major Goals}

\noindent$\bullet$ Study the solar (stellar) dynamo with rotation as the only important free parameter;

\noindent$\bullet$ investigate the energy transfer and heating of the stars' chromospheres, transition regions and coronae, as well as explosive events such as coronal mass ejections and flares;

\noindent$\bullet$ determine accurate rotation-age-activity relations for solar-type stars and, from these relations, the ages of G0--5~V stars can be estimated from knowing the rotation period, or values of coronal X-ray luminosity L$_{\rm x}$, Ca {\sc ii} $HK$ emission (or R'$_{HK}$), or other activity indicators;

\noindent$\bullet$ determine XUV irradiances of the young Sun to study the effects of its inferred high levels of XUV emissions and the resulting photo-ionization and heating of early planetary atmospheres;

\noindent$\bullet$ determine the ages and XUV flux histories of stars that host planets using rotation-age-activity relations and also to determine the effects of XUV radiation on these planets to determine if such planets could be suitable for life;

\noindent$\bullet$ continue to extend the sample to study the more common, lower mass and luminosity, cooler dK \& dM stars. These stars are very common, comprising about 90\% of all stars in the solar neighborhood.  Also an increasing number of planets are being found
hosted by dK/M stars.  Since the CZs of cooler stars are deeper (becoming fully convective around dM 3.5+), we can investigate the dynamo as a function of CZ depth by comparing dG, dK and dM stars having similar rotation periods and/or ages.

\section{Rotation Periods of the Program Stars}

Whenever possible, the rotation periods of the stars are directly measured from the rotation modulation in brightness arising from star spots (and bright faculae) and/or from rotational modulations of chromospheric Ca {\sc ii} $HK$ line emissions.  In some cases, UV chromospheric and transition region (TR) line emissions (such as Mg {\sc ii} $hk$ (2800\AA) and C {\sc iv} (1550\AA)) are used to determine the activity level and the rotation period -- such as for the G2~V star $\alpha$ Cen A (see Carton et al. 2007).  The determination of reliable rotation periods for younger stars (age $<$ 1.5 Gyr) is accomplished from conventional ($\pm$ $\sim$5 mmag precision) ground-based photometry. However, for older solar-type stars (ages $>$ 2 Gyr), the star spots typically cover less than 1\% of the stellar surface and rotational light modulations are difficult to detect. For example, the oldest solar-type star in our ongoing ground-based photometry program, found to show definite periodic light variations, is the G1~V star 15 Sge (age $\approx$ 1.9 Gyr; P $\approx$ 13.5 days; V-band light amplitude $\approx$ 0.008$\pm$0.003 mag). For comparison, the average rotationally-induced light modulation of the Sun, arising from bright plages and sunspots, is typically less than $\sim$0.15\% ($\sim$0.0015 mag). Although, with high precision photometry from space missions such as {\em MOST} (www.astro.ubc.ca/MOST/) and {\em CoRot} (smsc.cnes.fr/COROT/) it is now possible to detect low amplitude ($<$ 1 mmag) light variations.  And soon, with $Kepler$ (kepler.nasa.gov/), it will be easy to determine star spot modulated brightness variations as small as our Sun's, and thus rotation periods, and determine star spot areal coverage for solar-age and older stars. Fig. 2a shows the dependence of starspot-induced light variability (as a proxy for starspot areal coverage and magnetic activity) plotted against age.  As shown, the young solar-type stars have significant light variations of a few percent up to 10\% while the Sun and older stars have very small ($<$0.2\%) light variations.   

An excellent example of the high precision light curves that are now feasible from space missions is illustrated by the recent photometry of the young (age $\approx$ 0.7 Gyr) G5~V star $\kappa^1$ Cet using the Canadian microsatellite {\em MOST} -- See Walker et al. (2007). From this study, Walker et al. were able to determine a precise rotation period of P$_{\rm rot}$ = $8.77\pm0.04$ days (from the three MOST datasets combined: 2003, 2004 \& 2005), along with starspot coverage and differential rotation.  An example of the light curve obtained with {\em MOST} is shown in Fig. 2b. The light variations are modeled with the presence of two starspot regions very different in area and differential rotation. This high precision {\em MOST} light curve is illustrative of the thousands of even higher quality light curves expected from the {\em Kepler} mission.

\begin{figure}
\includegraphics[height=2in,width=5.3in,angle=0]{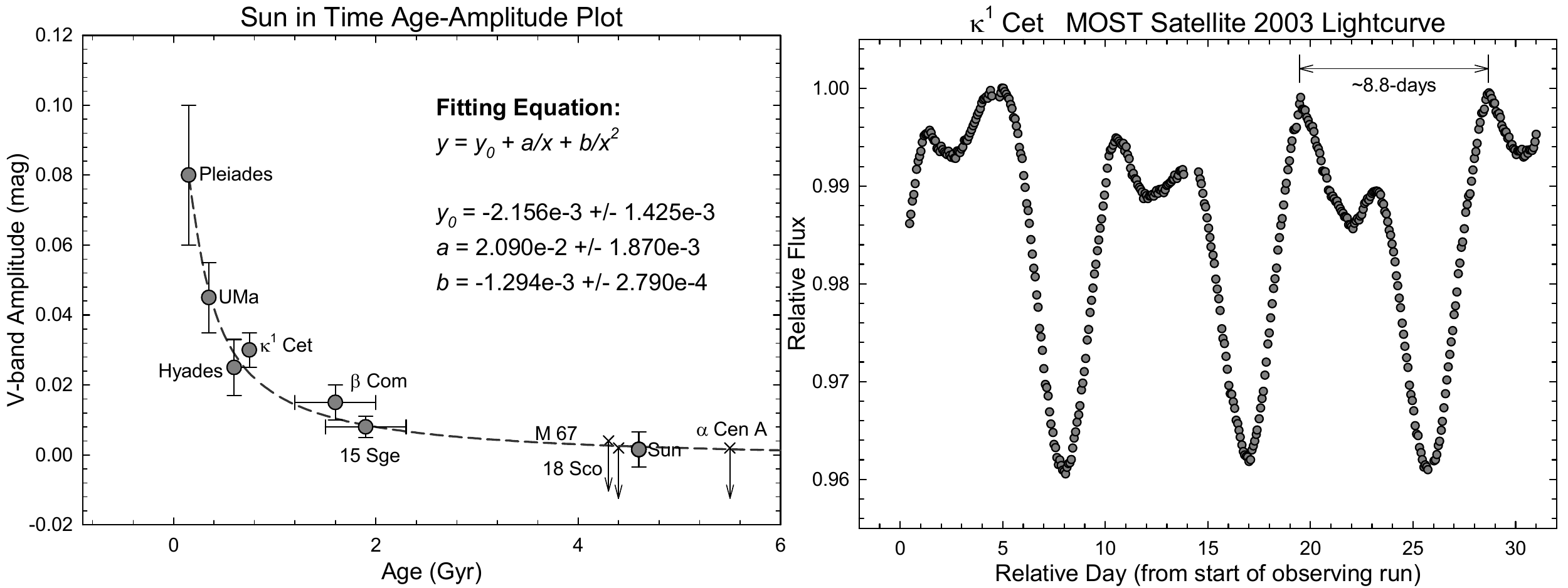}
\caption{{\bf Left (a) --} The V-band light amplitudes from starspot rotational modulations and cycles plotted against stellar age.  Note that the younger stars have significant light variations due to the larger areal coverage of starspots.  The least squares fitting equation for these data is given within the figure. {\bf Right (b) --} High precision photometry carried by Walker et al. (2007) with the {\em MOST} satellite of the Hyades-age G5 V star $\kappa^1$ Cet is shown. The $\sim$8.8 day brightness variations arise from the rotational modulation from at least two primary starspot regions with different areas on the surface of the star. }\label{fig:amp}
\end{figure}

Determinations of rotation periods also can be made from time-series measures of Ca~{\sc ii} $HK$ emission.  Most of these Ca {\sc ii} emission studies have been made at Mt. Wilson and more recently from Lowell Observatory (e.g. see Baliunas et al. 1995; Hall et al. 2007).  Although these programs are primarily focused on defining activity cycles, the period analyses also revealed, for some of the stars, apparent rotational modulations and rotation periods.  A number of our program stars have rotation periods found from Ca {\sc ii} $HK$ spectrophotometry.  Also, from time series UV spectrophotometry of chromospheric and TR emissions, the rotation periods of several more stars have been secured (see Datin et al. 2009; Guinan et al. 2009). 

As shown previously in Fig. 1a, the angular velocity ($\Omega$ $\propto$ 1/P$_{\rm rot}$) of our program stars decreases with age. By solar age, the light variations are $<$ 2-mmag and are barely detectable. The analytical expression giving the best fit to these data is provided with the plot. As shown in the figure, the rotation period of solar-type stars range from $<$ 3 days for the young (near-ZAMS) stars to $\gtrsim$ 35 days for the older 8--9 Gyr solar-type stars. Also shown is that, for a given age, the rotation period of a dM star is longer than the corresponding solar-type star. The more rapid magnetic braking of dM stars is most likely due to more efficient magnetic (and maybe turbulent) dynamos (from their deep CZs) as well as their lower masses. Not shown is that the initial rotation periods of very young near-ZAMS solar-type stars with ages less than 100 Myr (Pleiades age and younger) range from $\lesssim$ 1 day up to 3.5 days.  Excellent discussions about rotation and activity for near-ZAMS and pre-main-sequence stars in nearby star forming regions are given elsewhere in this volume.  The dispersion in initial rotation periods of these young stars arise from initial differences in the stars' angular velocities. Such differing conditions are -- stars having different initial angular velocities, interactions with planetary accretion disks and possible undetected binary companions.  However, by the age of about 300 Myr, the stars of a given age (for a set mass -- in this case one solar mass) have similar rotation periods.  For example, in the Hyades cluster (age = 650 Myr), most (if not all) of the single G0--5 V stars have rotation periods between 7.5 and 9 days.  This rotation ``funneling'' effect with age (large dispersion in the rotations for very young stars to low dispersion in rotation for older stars) probably results because the loss of angular momentum from magnetic braking is strongly dependent on the rotation of the star. Stars that initially are rapid rotators spin-down more rapidly from stronger magnetic braking than those that are originally slower rotators. If this self-regulating process were not active, we would not have rotation-age-activity relations.  Physically this is explicable by modern hydromagnetodynamo theory.  Some in-depth recent reviews of solar and stellar dynamo theory are given by Weiss and Tobias (2000) and Dikpati (2005), and most recently by Brandenburg (2009).

\section{Determining Reliable Ages for the Program Stars}

As shown by the review \& poster papers at this symposium, determining accurate stellar ages is crucial for astrophysics, but not an easy task.  When possible, the ages of program stars are estimated from memberships in nearby open clusters -- such as the Pleiades, Praesepe and Hyades clusters. Because there are few nearby solar-age stars with reliable ages, it would be important to use the $\sim$14--15 mag solar-type stars in M67 (age $\approx 4.3\pm0.5$ Gyr) to calibrate the rotation-age-activity relations.  Also these stars would be useful comparisons for the present Sun (i.e. serve as solar twins). At $\sim$850-pc, though, M67 is a bit too distant (and stars too faint) to reliably determine rotational periods or measure magnetic activity indicators (such as L$_{\rm x}$). However, Giampapa et al. (2006) have measured Ca {\sc ii} emissions for a number of the solar-type stars in M67 and find that the average Ca {\sc ii} emission level is similar to that of the Sun.  More recently, Pasquini et al. (2008) have identified several G2~V stars in M67 that nearly perfectly match the Sun in age, T$_{\rm eff}$, [Fe/H], Lithium abundance and age.  Unfortunately no time-series photometry or spectroscopic observations of the M67 cluster G0--5~V members have been carried out to determine their rotation periods.  This is feasible with a 2-m class telescope and, as part of the Sun in Time, we hope to carry out the high precision photometry during 2009/10. These observations would have to be carried out over a few months to cover their expected 20-30 day rotation periods and expected $<$ 2 mmag light variations if indeed these stars behave like the Sun. Also, Wright (2009 -- this volume) discusses the $\sim$2 Gyr ``benchmark'' open cluster -- Ruprecht 147.  This nearby ($\sim$200 pc) cluster would fill in the age gap in rotation-age-activity studies between the Hyades (650 Myr) and the Sun (4.57 Gyr).

Stellar ages are also estimated from UVW space motions for members identified with stellar moving groups -- such as the Pleiades, Castor, Ursa Major (UMa), Hyades and a few others. Soderblom et al. (1993) have intensively studied space motions and found many nearby dG and dK stars that are probable members of these moving groups. More recently, a few additional stars have been linked to the older ($\sim$2 Gyr) HR 1614 Moving Group (see Feltzing \& Holmberg 2000). For solar-age stars (and older) with well determined values of T$_{\rm eff}$, [Fe/H] and M$_{\rm V}$, ages can be estimated from isochronal fits using modern stellar evolution codes.  Several excellent codes (and isochrones) are currently available (e.g. VandenBerg et al. 2006; Yonsei-Yale Isochrones -- Demarque et al. 2004).  Unfortunately, isochronal fits for young stars (i.e. still close to the ZAMS) are not very reliable because of the sensitivity of age determinations to errors in mass, T$_{\rm eff}$, $\log~g$, M$_{\rm V}$ and metal abundance. It should be noted that the stars are selected either from spectral types (G0--G5~V) or corresponding $B-V$ color indices (+$0.60-0.70$). However, it is preferable to use Str\"omgren photometry ($b-y$) instead of Johnson/Bessell ($B-V$) to estimate T$_{\rm eff}$ since the ($b-y$) index is less sensitive to the metallicity of the star.

\section{Magnetic Activity Indicators -- Coronal X-ray, Transition Region \& Chromospheric FUV--UV Emissions and Starspots}

\subsection{Coronal X-ray and FUV/UV Transition Region \& Chromospheric Emissions}

Almost all of the primary program stars have magnetic activity indicators since this was a major factor in the selection process.  In particular, because this study was initiated when the International Ultraviolet Explorer ({\em IUE}) satellite was operating (1978--1995), most of the stars have {\em IUE} UV spectra.  The wavelength interval covered by {\em IUE} (1160--3200\AA) contains numerous important chromospheric and TR emission lines that include Ly-$\alpha$ 1215.6\AA, Mg {\sc ii} $h$ \& $k$ 2800\AA~and strong TR lines such as C {\sc iii} 1335\AA, Si {\sc iv} 1400\AA, C {\sc iv} 1550\AA, He {\sc ii} 1640\AA~and others. Also, during the 1990s, {\em ROSAT} (0.1--2.5 keV) and {\em ASCA} (0.2--10 keV) were used to secure pointed X-ray observations of several young, active program stars.  These observations are supplemented by archival {\em ROSAT} (both pointed and all sky survey data) and {\em ASCA} X-ray observations.  Archival X-ray observations obtained more recently by {\em Chandra} and {\em XMM-Newton} are also used.  Observations were also carried out in the 80--500\AA~wavelength range with the Extreme Ultraviolet Explorer ({\em EUVE}). The EUV region contains important diagnostic coronal and TR emissions lines.  Because {\em EUVE} had a small effective collection area, only a few solar-type stars were observed. In addition to these data, FUV spectra of solar-type stars -- particularly high dispersion spectrophotometry of the strong H {\sc i} Ly-$\alpha$ feature -- were also used in the FUV irradiance studies.  Ly-$\alpha$ emission is the largest contributor to the X-FUV (1--1700\AA) flux, comprising 70--90\% of the total flux, in that wavelength range, for solar-type stars.  Wood et al. (2002, 2005) use {\em HST} spectroscopy of the Ly-$\alpha$ feature to estimate the winds and mass loss rates of solar-type (and cooler) stars. They provide Ly-$\alpha$ emission fluxes for the stars in their sample, after the careful removal of interstellar absorption and geocoronal emission.

\subsection{Chromospheric Ca {\sc ii} H \& K emissions}

Most of the program stars have measures of chromospheric Ca {\sc ii} $HK$ emission. Almost all of these observations are from Mount Wilson or Lowell Observatory. The Ca {\sc ii} $HK$ emissions are important, widely available and well-calibrated indicators of chromospheric activity and numerous papers have focused on these lines. (See e.g. Soderblom et al. 1993; Baliunas et al. 1995; Hall et al. 2007; Mamajek (in this volume); and references therein). One drawback with using Ca {\sc ii} emission is that it levels off for stars older than $\sim$3--4 Gyr. Thus it becomes insensitive to rotation and age for older stars.  Also, care must be taken in using Ca {\sc ii} $HK$ emission measurements to determine age and rotation because chromospheric emission can vary from rotation modulation, flares, and also can have relatively large cyclic variations. Multiple measures should be used.

\subsection{Light modulations by starspots} 

A usually overlooked indicator of activity is starspot areal coverage which is manifested by low amplitude brightness variations. Young, chromospherically-active stars have relatively large rotationally modulated, periodic brightness variations arising from an uneven distribution of cool starspots over the stars' surfaces.  For example, Pleiades G0--5~V stars have light variations of typically $\sim$0.06--0.12 mag whereas the Sun has light modulations of $\sim$0.12\% (0.0012 mag = 1.2 mmag). Recently, from long-term high precision photometry, Hall et al. (2007) report brightness variation as small as $\approx$ 0.1 percent (1~mmag) for the solar twin -- 18 Sco (G2~V; age $\sim$ 4.4 Gyr).  Although the light amplitudes change from season to season (from activity cycles and differential rotation), the light modulations and resulting rotation periods could make nice age indicators, as shown in Fig. 2a. Excellent long-term high precision photometry and Ca {\sc ii} $HK$ spectrophotometry of solar-type stars have been carried out at Lowell Observatory (See e.g. Lockwood et al. 2007).  Results from our long-term photometry of some program stars are given by Messina et al. (2006) and references therein.

\section{Effects of the Young Active Sun's XUV \& Plasma Fluxes on Terrestrial Planets}

The Sun in Time program also focuses on the crucial question of the influence of the young Sun's strong XUV emissions on the developing planetary system -- in particular on the photo-chemical and photo-ionization evolution and possible erosion of their early atmospheres. The XUV spectral irradiance (1--1700\AA) measures have been made for stars covering ages from $\sim$100 Myr to 6.7 Gyr (Ribas et al. 2005). These data are of interest to researchers of young planetary atmospheres and for studies of the atmospheric evolution of the large number of extrasolar planets found orbiting other solar-type stars. As an illustration, Fig. 3a shows a plot of the FUV O {\sc vi} irradiances (i.e.-- flux densities at 1 AU) for solar-type stars of different ages. The emission fluxes decrease by nearly 50$\times$ over the stellar ages sampled. Fig. 3b shows the averaged XUV irradiance over time, inferred for the Sun and other solar-type (dG0--5) stars, from this study. The fluxes in the plot have been normalized to the corresponding mean flux values of the present Sun. As shown, the coronal X-ray/EUV emissions of the young main sequence Sun were $\sim100-1000\times$ stronger than the present Sun. Similarly, the TR and chromospheric FUV \& UV emissions of the young Sun (by inference) were $10-100\times$ and $5-10\times$ stronger, respectively, than at present. Also shown in the figure is the slow increase ($\sim$8\%/Gyr) in the bolometric luminosity of the Sun over the past 4.6 Gyr. Over this time, the Sun's luminosity increased from $\sim$0.7 L$_\odot$ to 1.0 L$_\odot$. So in the past, the young Sun was dimmer by 30\% overall but, due to strong magnetic activity, had stronger XUV fluxes than today.

\begin{figure}
\includegraphics[height=2.7in,width=5.3in,angle=0]{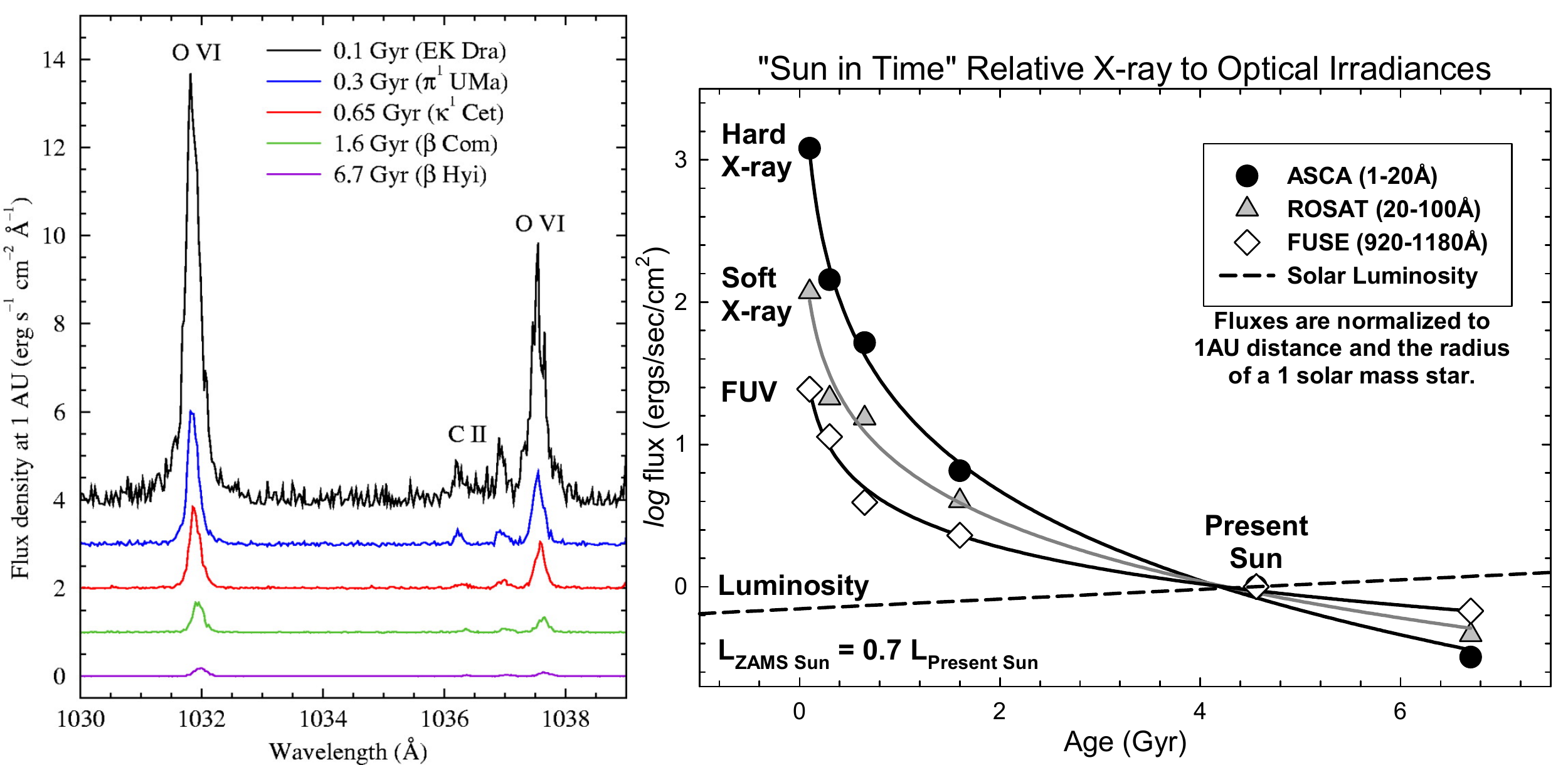}
\caption{{\bf Left (a) --} The FUV irradiance of the O {\sc vi} spectral region of six ``Sun in Time'' program stars with ages from $\sim$130 Myr to 9 Gyr. As shown, the emission fluxes decrease by nearly 50$\times$ from the youngest sun to the oldest sun in the sample. {\bf Right (b) --} Irradiances (flux densities at a reference distance of 1~AU) of the Sun in Time program stars are plotted for hard (1--20\AA) \& soft (20--100\AA) X-rays (from {\em ASCA} and {\em ROSAT}, respectively) along with the total FUV emission fluxes (from FUSE). The fluxes have been normalized to the corresponding mean fluxes of the Sun. These measures are based on the paper by Ribas et al. (2005). Also plotted is the change in the bolometric luminosity (L) of the Sun based on modern stellar evolution models. Note that the ZAMS Sun had a luminosity of 30\% less than today.}\label{fig:kappa}
\end{figure}

Since 2002 we have been collaborating with planetary scientists and astrobiology groups (in particular -- Helmut Lammer and others at the Astrobiology Group at the Space Research Institute, Graz, Austria) to study the effects of the strong XUV radiation and inferred high plasma fluxes of the young Sun on hosted planets.  Brief summaries of some of the results are discussed below for Mercury, Venus and Mars, as well as for exoplanets.  Because of space constraints, we omit a discussion of the early Earth except to say that the Earth was shielded to a large extent from the sputtering / ion pickup processes discussed below because of the its protective magnetic field.  A good review of the early evolution of the Earth's atmosphere (and the effects of the young Sun's high XUV and plasma fluxes) is given by Lammer et al. (2008). 

\subsection{Possible Erosion of Mercury by the Active Young Sun} 
Mercury is the nearest planet to the Sun at a distance of 0.39 AU.  Because of its proximity to the Sun, Mercury is subjected to over 6.5$\times$ more solar radiation and wind fluxes than the Earth.  Thus, early Mercury was much more strongly exposed to the early Sun's very high particle fluxes and XUV radiation than any other planet.  Even today, with a kinder, more gentle Sun, the solar wind and coronal \& chromospheric XUV emissions (as well as flares and coronal mass ejection events) produce sputter/ion pickup erosion of its surface in which ions of Na$^+$, K$^+$, and O$^+$ and even OH$^+$ and H$_2$O$^+$ have been observed in its very thin, transient exosphere. (See e.g. Zurbuchen et al. 2008.)

As pointed out by Tehrany et al. (2002), some of the unusual features of Mercury -- such as its high mean density (5.43 gm/cm$^3$) for its size -- can be explained by the extreme erosive XUV/plasma actions of the early Sun.  Mercury is an exceptionally dense planet for its size because of the large relative size of its core.  Mercury's iron core extends out to 74\% of its radius and it is estimated that the iron core occupies about 42\% of its volume; for Earth this proportion is 17\%.  Because of this, Mercury is often referred to as the ``Iron Planet''.  Preliminary modeling of early Mercury by Tehrany et al. (2002) was carried out using solar XUV irradiances from the Sun in Time program and a range of plasma fluxes. This study indicates that Mercury could have undergone significant erosion (by sputtering and ion pickup mechanisms) of its surface and mantle during the Sun's active phase -- the first $\sim$0.5--1.0 Gyr of its lifetime.  This hypothesis is speculative and there are several other explanations; one of these is a major collision with large planetesimal that may have stripped away Mercury's outer mantle.  Perhaps this question will be answered with measurements being carried out by the Mercury mission MESSENGER (www.nasa.gov/messenger) and also by BepiColumbo (www.esa.int/science/bepicolombo).

\subsection{The Loss of the water inventory from Venus}
Using the XUV irradiance appropriate for the young Sun (from Ribas et al. 2005) and solar wind flux estimates from Wood et al. (2002, 2005), Kulikov et al. (2006) have studied the atmosphere and water loss from early Venus.  Venus has D/H ratios indicating that the juvenile planet once had water, possible even oceans.  But Kulikov et al. show that Venus (at 0.71 AU from the Sun) has essentially lost all of its water reserves during the first $\sim$0.5 Gyr after its formation from the vigorous action of strong (massive) winds and high XUV fluxes.  However, as discussed recently by Lammer et al. (2008) and others, there are many uncertainties about early Venus that still leave many unanswered questions and problems.  For example, Venus today rotates very slowly and does not have a protective magnetic field. But did early Venus rotate faster and have a strong dynamo with a resulting protective magnetosphere?  Because of these and other uncertainties about Venus' early properties and initial conditions, many questions about its early history remain.  

\subsection{Loss of Water and soil oxidation of Mars} 
It has been assumed, from topographic and geological studies, that Mars was originally warmer and much wetter than at present, and likely possessed a $\sim$1 bar atmosphere (see discussions in Lammer et al. 2008).  Lammer et al. (2003) considered ion pick-up sputtering as well as dissociative recombination processes to model the effects of the active young Sun's high XUV \& plasma fluxes on the planet.  The loss of H$_2$O from Mars over the last 3.5 Gyr was estimated to be equivalent to a global Martian H$_2$O ocean with a depth of $\gtrsim$12-m.  If ion momentum transport is important on Mars, the water loss may be enhanced by a factor of $\sim$2.  For their study it has been assumed that, for the first billion years, Mars had a hot liquid iron-nickel core and, through rotation, possessed a significant magnetic field and resulting magnetosphere.  This magnetosphere essentially shielded the early Martian environment from the combination of high levels of solar XUV radiation and plasma fluxes that would have otherwise removed its atmosphere.  However, Mars is a smaller, less massive planet than the Earth [M$_{\rm Mars}$ $\approx$ 1/10 M$_\oplus$] with a smaller iron core which lost heat at a much faster rate.  Thus its iron core is expected to have solidified $\sim$1 billion years after the planet's formation.  Without the protective magnetic field, the Martian exosphere was exposed to the ionizing effects and strong winds of the early Sun, and thus partially eroded.  Photolysis of water (H$_2$O $\rightarrow$ 2H + O) ensued, with a preferential loss of the lighter hydrogen over the (8$\times$) heavier oxygen.  The loss of water and water vapor from the atmosphere resulted in a greatly diminished greenhouse effect and subsequent rapid cooling of the lower Martian atmosphere.  This rapid cooling permitted some water to remain behind, possibly as ice and permafrost trapped below the Martian surface. 

\subsection{Exoplanets -- Possible Evaporation of Hot Jupiters} 
Our XUV irradiance data from Ribas et al. (2005) have been used by Grie{\ss}meier et al. (2004) to investigate the atmospheric loss of extrasolar planets from the XUV heating and winds of their host stars. This study indicates (among other interesting results) that strong planetary magnetic fields and resulting significant magnetospheres are critically important in shielding a planet from the atmospheric erosion and possible atmospheric loss (via sputtering and ion pickup mechanisms) expected from the strong winds and high XUV radiation of the active young Sun.  This is particularly important for lower mass, terrestrial size planets with low gravities, located close to their host stars. 

\section{The Next Step:  Extension of the ``Sun in Time'' Program to Main-sequence K- and M- Stars}

The discovery of 330+ extrasolar planets orbiting mostly main-sequence dG, dK, and dM stars during the last decade has motivated the expansion of the ``Sun in Time'' program to include dK \& dM stars.  These cool, low mass and low luminosity stars comprise over 90\% of all stars in the Galaxy.  With the numerous ground-based extrasolar planet search programs and space missions such as {\em Kepler}, thousands of additional extrasolar planets are expected to be found in the near future.  The goals of our dK--dM star program are to understand magnetic activity of red dwarf stars with deep CZs and to determine their XUV and plasma fluxes over a wide range of ages.  This program is helping to identify and characterize dK \& dM stars that have planets which may be suitable for life. Many dK \& dM stars are members of the old disk and Pop II populations of our Galaxy. In particular, late dK \& dM stars make interesting targets for exobiology and SETI programs because of their long lifetimes, exceeding 50 Gyr, and high space frequencies; See Tarter et al. (2007).  

Because of the low luminosities of dM stars (e.g. L $\approx$ 0.06 L$_\odot$ for dM0 stars and L $\approx$ 0.008 L$_\odot$ for dM5 stars), their liquid water habitable zones (HZs) are located close to the host star ($\lesssim$0.4 AU).  However, because of their deep CZs and efficient magnetic dynamos, late dK \& dM red dwarf stars exhibit strong XUV coronal and chromospheric emissions and frequent flares.  The high magnetic activity of dM stars causes a hypothetical hosted HZ planet to be strongly affected by stellar flares, winds and plasma ejection events that are frequent in young dM stars (e.g. Kasting et al. 1993; Lammer et al. 2006; Guinan \& Engle 2007).  Relevant to the ``Ages of Stars'' Symposium, we are developing rotation-age-activity relations for these ubiquitous little stars.  Because of the very slow nuclear evolution of dK \& dM stars, applying isochronal fits to determine age is not feasible.  Reliable ages can be obtained from cluster and moving group memberships, as well as memberships in wide binary systems (such as Proxima Cen (M5.5~V) as an outlying member of the $\alpha$ Cen triple star system) in which the age of the companion is determinable from isochrones or, in the case of white dwarf companions, from the white dwarf cooling ages -- See Catalan (2009) in this volume.   

A description of this program -- ``Living with a Red Dwarf'' -- can be found on our website at (www.astronomy.villanova.edu/lward/). Also, some recent papers discussing initial results are given by Guinan and Engle (2009) \& Engle et al. (2009).

\begin{acknowledgments}
This research is supported by grants from NSF and NASA and utilizes data from the IUE, ROSAT, ASCA, EUVE, FUSE, HST, XMM-Newton \& the Chandra missions.  The photometry used in the program is supported by a NSF/RUI grant.  We are very grateful for this support. We also thank the Symposium organizer, David Soderblom, for the invitation to participate.
\end{acknowledgments}

\end{document}